\documentstyle[12pt,epsfig]{article}
\def\np#1#2#3  {{\it Nucl. Phys. }{\bf #1} (19#3) #2} 
\def\nc#1#2#3  {{\it Nuovo. Cim. }{\bf #1} (19#3) #2} 
\def\pl#1#2#3  {{\it Phys. Lett. }{\bf #1} (19#3) #2} 
\def\pr#1#2#3  {{\it Phys. Rev. }{\bf #1} (19#3) #2} 
\def\prl#1#2#3  {{\it Phys. Rev. Lett. }{\bf #1} (19#3) #2} 
\def\prep#1#2#3 {{\it Phys. Rep. }{\bf #1} (19#3) #2} 
\def\zp#1#2#3  {{\it Z. Phys. }{\bf #1} (19#3) #2} 
\def\rmp#1#2#3  {{\it Rev. Mod. Phys. }{\bf #1} (19#3) #2} 
\def\hepph  #1 {{\tt hep-ph/#1}}

\begin{document}

\def\tr{\mathop{\rm tr}}
\def\Tr{\mathop{\rm Tr}}
\def\Im{\mathop{\rm Im}}
\def\Re{\mathop{\rm Re}}
\def\bR{\mathop{\bf R}{}}
\def\bC{\mathop{\bf C}{}}
\def\C{\mathop{\rm C}}
\def\bra#1{\left\langle #1\right|}
\def\ket#1{\left| #1\right\rangle}
\def\VEV#1{\left\langle #1\right\rangle}
\def\gdot#1{\rlap{$#1$}/}
\def\abs#1{\left| #1\right|}
  \newcommand{\ccaption}[2]{
    \begin{center}
    \parbox{0.85\textwidth}{
      \caption[#1]{\small{\it{#2}}}
      }
    \end{center}
    }
\def\beq{\begin{equation}}
\def\eeq{\end{equation}}
\def\eq{\beq\eeq}
\def\beqn{\begin{eqnarray}}
\def\eeqn{\end{eqnarray}}
\relax
\let\h=\hat
\newcommand\SS{\scriptsize}
\newcommand\sss{\scriptscriptstyle}
\newcommand\gs{g_{\sss S}}
\newcommand\as{\alpha_{\sss S}}         
\newcommand\ep{\epsilon}
\newcommand\Th{\theta}
\newcommand\epb{\overline{\epsilon}}
\newcommand\aem{\alpha_{\rm em}}
\newcommand\refq[1]{$^{[#1]}$}
\newcommand\avr[1]{\left\langle #1 \right\rangle}
\newcommand\lambdamsb{\Lambda_5^{\rm \sss \overline{MS}}}
\newcommand\qqb{{q\bar{q}}}
\newcommand\qb{\bar{q}}
\newcommand\xto{\tilde{x}_1}
\newcommand\xtt{\tilde{x}_2}
\newcommand\aoat{a_1 a_2}
\newcommand\Oop{{\cal O}}
\newcommand\Sfun{{\cal S}}
\newcommand\Pfun{{\cal P}}
\newcommand\mug{\mu_\gamma}
\newcommand\mue{\mu_e}
\newcommand\muf{\mu_{\sss F}}
\newcommand\mufp{\mu_{\sss F}^\prime}
\newcommand\mufs{\mu_{\sss F}^{\prime\prime}}
\newcommand\mur{\mu_{\sss R}}
\newcommand\murp{\mu_{\sss R}^\prime}
\newcommand\murs{\mu_{\sss R}^{\prime\prime}}
\newcommand\muh{\mu_{\sss H}}
\newcommand\muhp{\mu_{\sss H}^\prime}
\newcommand\muhs{\mu_{\sss H}^{\prime\prime}}
\newcommand\muo{\mu_0}
\newcommand\MSB{{\rm \overline{MS}}}
\newcommand\DIG{{\rm DIS}_\gamma}
\newcommand\CA{C_{\sss A}}
\newcommand\DA{D_{\sss A}}
\newcommand\CF{C_{\sss F}}
\newcommand\TF{T_{\sss F}}
\newcommand\pt{p_{\sss T}}
\newcommand\kt{k_{\sss T}}

\begin{titlepage}
\nopagebreak
{\flushright{
        \begin{minipage}{4cm}
        CERN-TH/99-91 \hfill \\
        ETH-TH/99-08 \hfill \\
        hep-ph/9904320\hfill \\
        \end{minipage}        }

}
\vfill
\begin{center}

{\large {\sc Jet cross sections in polarized\\
             photon--hadron collisions}} 
\vskip .5cm
{\bf Daniel de Florian}
\\                    
\vskip .1cm
{Theoretical Physics, ETH, Zurich, Switzerland} \\
\vskip .5cm
{\bf Stefano Frixione}\\                    
\vskip .1cm
{CERN, TH Division, Geneva, Switzerland}\\
\end{center}
\nopagebreak
\vfill
\begin{abstract}
We present a computation of one- and two-jet cross sections in
polarized photon--hadron collisions, which is accurate to
next-to-leading order in QCD. Our results can be used to compute
photoproduction cross sections in electron--proton scattering. To
this purpose, we investigate the structure of the polarized
Weizs\"acker--Williams function, where we include a universal,
non-logarithmic term, neglected in the literature. We construct a
Monte Carlo code, within the framework of the subtraction method,
and we use it to study the phenomenology of jet production in the
energy range relevant to HERA. In particular, we investigate the
perturbative stability of our results, and we discuss the possibility 
of constraining polarized parton densities of the proton and the photon 
using jet data.
 
\end{abstract}        
\vfill
CERN-TH/99-91 \hfill \\
April 1999\hfill
\end{titlepage}

The last decade has seen an important advance in our understanding of
polarized nucleon structure functions as a result of the analysis of 
deep inelastic scattering (DIS) data~\cite{data}. Unfortunately, the 
use of DIS data alone does not allow an accurate 
determination of the polarized parton densities.
This is true in particular for the gluon, since this quantity
contributes to DIS in leading order (LO) only via the $Q^2$-dependence 
of the spin asymmetry ($A_1^N$), which could not be thoroughly studied
experimentally so far. In the near future, the improved capabilities
of handling polarized beams will allow high-energy colliders to
operate in polarized mode, thus giving the possibility to study
polarized scattering in a way complementary to DIS. At the RHIC
collider at BNL, the first proton--proton polarized collisions 
are expected in two years from now. The possibility that the 
electron--proton HERA collider at DESY will run with polarized 
beams has also been considered for a long time~\cite{HERApol}.

At variance with DIS, collider physics offers a relatively large 
number of processes whose dependence upon the gluon density is
dominant already at LO. The study of these processes is therefore
crucial in order to measure this density in a direct way.
Among the various high transverse momentum reactions with a
sizeable gluonic contribution, jet production is an obvious
candidate, because of the large rates. At HERA, the largest
cross sections will be obtained by retaining those events
where the electron is scattered at very small angles with respect to
the beam line. In this way, the photon exchanged between the electron
and the proton is almost on-shell, and the underlying dynamics is
that of a photoproduction process.

As was shown in ref.~\cite{svhera}, a polarized version of the HERA
collider with $\sqrt{S}\approx 300$ GeV would be a very promising and
useful facility for studying polarized photoproduction reactions. In
particular, jet production cross sections display a strong sensitivity to
the polarized gluon distribution of the proton. However, the analysis
of ref.~\cite{svhera} is based on a lowest order QCD computation.
From unpolarized physics it is well known that, in order to describe 
the data in a satisfactory way, the calculation of next-to-leading order 
(NLO) QCD corrections to the jet cross sections is mandatory.

The computation of the NLO QCD corrections to one- and two-jet production 
in polarized photon--hadron collisions is the purpose of this paper. As is
well known, in QCD photoproduction cross sections are written as the
sum of two terms (point-like and hadronic components), neither of which is
physically meaningful by itself. The hadronic part, in which the 
photon behaves like a hadron, can be accounted for by using the 
available results for jet production in polarized hadron--hadron
scattering~\cite{FFSV}. The NLO QCD corrections for the point-like 
part are presented in this paper for the first time.
Such a calculation needs the one-loop $2\to 2$ and tree-level $2\to 3$
polarized amplitudes as input (one of the particles in the initial state 
being the photon and the other a quark or gluon). These amplitudes 
are already known~\cite{dfw}. The next step is to implement them in a 
Monte Carlo code which allows the calculation of any three-parton
infrared-safe observable. In order to do this, we will use the 
general algorithm, based on the subtraction method, presented
in ref.~\cite{FKS}. Following the same strategy as already adopted 
in the case of polarized hadron--hadron collisions~\cite{FFSV}, we
build a modified version of the Monte Carlo code of ref.~\cite{frixione}, 
which deals with unpolarized photon--hadron collisions. The reader can
find further details in ref.~\cite{FFSV}.

In photoproduction processes initiated by an electron, the electron 
can be considered to be equivalent to a beam of real photons, whose 
distribution in energy (Weizs\"acker--Williams function~\cite{wwori}) 
can be computed theoretically. The polarized electron--hadron
cross section reads
\beq
d\Delta\sigma_{eH}(K_e,K_H)=\int_{y_{min}}^{y_{max}}
dy \,\Delta f_\gamma^{(e)}(y)\,d\Delta\sigma_{\gamma H}(yK_e,K_H),
\label{sigeppol}
\eeq
where $K_e$ and $K_H$ are the momenta of the incoming electron and
hadron respectively, and \mbox{$0<y_{min}<y_{max}\leq 1$} are fixed by
kinematical boundary conditions or experimental cuts. The quantities
$\Delta\sigma$ appearing in eq.~(\ref{sigeppol}) are defined in terms
of cross sections \mbox{$\sigma(\lambda_1,\lambda_2)$} for incoming 
particles of definite helicities
\beq
d\Delta\sigma=
\frac{1}{4}\Big(d\sigma(+,+)+d\sigma(-,-) 
-\,d\sigma(+,-)-d\sigma(-,+)\Big).
\label{Deltadef}
\eeq
The unpolarized electron--hadron cross section can be obtained
from eq.~(\ref{sigeppol}) with the formal substitutions
\beq
\Delta\sigma\longrightarrow\sigma,\;\;\;\;\;\;
\Delta f_\gamma^{(e)}\longrightarrow f_\gamma^{(e)}.
\eeq
As mentioned before, the polarized and unpolarized Weizs\"acker--Williams
functions (\mbox{$\Delta f_\gamma^{(e)}$} and \mbox{$f_\gamma^{(e)}$},
respectively) can be computed. As far as the unpolarized case
is concerned, we use the form~\cite{wwfmnr}
\beq
f_{\gamma}^{(e)}(y)=\frac{\aem}{2\pi}
\left[
\frac{1+(1-y)^2}{y}\log\frac{Q^2(1-y)}{m_e^2 y^2}
+2 m_e^2 y\left(\frac{1}{Q^2}-\frac{1-y}{m_e^2 y^2}\right)\right].
\label{wwfununp}
\eeq
Notice that eq.~(\ref{wwfununp}) contains a non-logarithmic term with a
singular behaviour for \mbox{$y\to 0$}. In ref.~\cite{sg2} this
term was shown to give non-negligible contributions (of about
7\% and 5\% for $Q^2=0.01$~GeV$^2$ and $Q^2=4$~GeV$^2$, respectively)
in the case of unpolarized jet production at HERA. One expects an analogous 
non-logarithmic term to appear also in the polarized Weizs\"acker--Williams 
function. To the best of our knowledge, however, such a term has never
been presented in the literature. We therefore computed 
\mbox{$\Delta f_\gamma^{(e)}$} from scratch. We closely follow the
procedure of ref.~\cite{wwfmnr}, in which the interested reader 
will find further details. We consider the generic production process
\beq
e(p;n)+a(k;s)\,\longrightarrow\,e(p^\prime)+X,\;\;\;\;
q=p-p^\prime,
\eeq
where $a$ is a parton whose momentum $k$, which is kept off-shell 
for the time being, will eventually be put on-shell, and $n$, $s$
are spin vectors. In the case of polarized scattering, we only need 
the antisymmetric part of the partonic and leptonic tensors
(see for example ref.~\cite{gr}):
\beqn
&&W^{\mu\nu}_A=
\frac{W^{\mu\nu}-W^{\nu\mu}}{2}=
i\frac{\sqrt{\abs{k^2}}}{k\cdot q}
\epsilon^{\mu\nu\rho\sigma}q_\rho
\left[g_1(q^2,k\cdot q) s_\sigma + 
g_2(q^2,k\cdot q)
\left(s_\sigma-\frac{q\cdot s}{q\cdot k}k_\sigma\right)\right],
\phantom{aa}
\label{parttens}
\\
&&T^{\mu\nu}_A=
\frac{T^{\mu\nu}-T^{\nu\mu}}{2}=
-2im_e \epsilon^{\mu\nu\rho\sigma}n_\rho q_\sigma,
\eeqn
where $g_1$ and $g_2$ are the usual structure functions relevant
to polarized collisions. The spin vectors for the incoming electron 
and parton can be written as 
\beq
n_\mu=N_n \left(p_\mu-\frac{m_e^2}{k\cdot p}k_\mu\right),\;\;\;\;
s_\mu=N_s \left(k_\mu-\frac{k^2}{k\cdot p}p_\mu\right),
\eeq
where $N_n$ and $N_s$ are normalization factors such that
\mbox{$\abs{n^2}=\abs{s^2}=1$}. The spin vectors satisfy the
transversality conditions \mbox{$n\cdot p=0$} and \mbox{$s\cdot k=0$}. 
The electron--parton cross section follows from direct 
computation\footnote{We keep only the first term in the expansion
of $g_1$ and $g_2$ in series of $q^2$: the remaining terms give 
non-factorizable corrections which are suppressed by powers of 
$Q^2/M_X^2$.} ($y=(k\cdot q)/(k\cdot p)$, $k^2\to 0$):
\beqn
d\Delta\sigma_{ea}(p,k)&=&
\frac{1}{4k\cdot p}\frac{e^2 W^{\mu\nu}_A T_{A\mu\nu}}{q^4}
\frac{d^3p^\prime}{(2\pi)^2 2E^\prime}
\nonumber \\
&=&-\frac{\aem}{2\pi}\left[\frac{1-(1-y)^2}{y q^2}
+\frac{2m_e^2y^2}{q^4}\right] \frac{g_1(0,k\cdot q)}{4y\,k\cdot p} dq^2 dy.
\label{sigmaea}
\eeqn
\enlargethispage*{50pt}
Notice that the terms proportional to $g_2$ drop from this equation.
This expression can be related to the polarized cross section for
real photon--parton scattering, which can be computed by convoluting
the parton tensor of eq.~(\ref{parttens}) with the antisymmetric part 
of the photon polarization density matrix, 
\beq
P^{\mu\nu}_A=\frac{1}{2}\left(\varepsilon^\mu {\varepsilon^\nu}^\star
- \varepsilon^\nu {\varepsilon^\mu}^\star\right)
=\frac{i}{2\sqrt{\abs{q^2}}}\epsilon^{\mu\nu\rho\sigma}
q_\rho t_\sigma,
\eeq
where $\varepsilon^\mu$ is the photon polarization vector and 
$t^\mu$ is its spin vector
\beq
t_\mu=N_t \left(q_\mu-\frac{q^2}{k\cdot q}k_\mu\right).
\eeq
We get ($q^2\to 0$)
\beq
\Delta\sigma_{\gamma a}(q,k)=\frac{1}{4k\cdot q} W^{\mu\nu}_A P_{A\mu\nu}=
\frac{g_1(0,k\cdot q)}{4k\cdot q}.
\label{sigmagammaa}
\eeq
Combining eqs.~(\ref{sigmaea}) and~(\ref{sigmagammaa}),
and integrating over $q^2$ in the accessible kinematical range
(see ref.~\cite{wwfmnr}), we get
\beq
d\Delta\sigma_{ea}(p,k)=
\Delta\sigma_{\gamma a}(yp,k)\,\Delta f_{\gamma}^{(e)}(y)\,dy,
\eeq
where
\beq
\Delta f_{\gamma}^{(e)}(y)=\frac{\aem}{2\pi}
\left[\frac{1-(1-y)^2}{y}\log\frac{Q^2(1-y)}{m_e^2 y^2}
+2 m_e^2 y^2\left(\frac{1}{Q^2}-\frac{1-y}{m_e^2 y^2}\right)\right].
\label{wwfunpol}
\eeq
We therefore find that also in the case of polarized scattering a
non-logarithmic term is present. This term, at variance with the
unpolarized case, is not singular for $y\to 0$. However, exactly
as in the unpolarized case, its behaviour for small $y$ is equal
to the behaviour of the term that multiplies the logarithm.

We can now go back to eq.~(\ref{sigeppol}), and compute
\mbox{$d\Delta\sigma_{\gamma H}$}. Using the factorization
theorems of QCD, we can write
\beq
d\Delta\sigma_{\gamma H}(K_\gamma,K_H)=
d\Delta\sigma^{\rm point}_{\gamma H}(K_\gamma,K_H)
+d\Delta\sigma^{\rm hadr}_{\gamma H}(K_\gamma,K_H),
\label{sigmagp}
\eeq
where
\beqn
d\Delta\sigma^{\rm point}_{\gamma H}(K_\gamma,K_H)&=&\sum_j\int dx
\Delta f^{(H)}_j(x,\muhp)
d\Delta\hat{\sigma}_{\gamma j}(K_\gamma,xK_H,\as(\murp),\murp,\muhp,\mug),
\label{pointcomp}
\\
d\Delta\sigma^{\rm hadr}_{\gamma H}(K_\gamma,K_H)
&=&\sum_{ij}\int dx dy
\Delta f^{(\gamma)}_i(x,\mug) \Delta f^{(H)}_j(y,\muhs)
\nonumber \\*&&\phantom{\sum_{ij}\int dx}\times
d\Delta\hat{\sigma}_{ij}(xK_\gamma,yK_H,
\as(\murs),\murs,\muhs,\mug)\,.
\label{hadrcomp}
\eeqn
We remind the reader that the two terms in the RHS of eq.~(\ref{sigmagp})
(denoted as point-like and hadronic components) are not separately well 
defined in perturbative QCD beyond leading order: only their sum 
is physically meaningful. The polarized parton distribution
functions \mbox{$\Delta f^{(H)}_i$} and \mbox{$\Delta f^{(\gamma)}_i$}
are defined, as usual, in terms of densities for partons of definite
helicity in hadrons of definite helicity ($A=H,\gamma)$:
\beq
\Delta f^{(A)}_i=f^{(A+)}_{i+}-f^{(A+)}_{i-}
=f^{(A-)}_{i-}-f^{(A-)}_{i+}.
\label{Deltafdef}
\eeq
The subtracted partonic cross sections \mbox{$d\Delta\hat{\sigma}_{\gamma j}$}
and \mbox{$d\Delta\hat{\sigma}_{ij}$} are finite at any order in QCD and, 
for the purpose of this paper, have been evaluated to NLO ($\aem\as^2$
and $\as^3$, respectively).  Notice that while the dependence upon the
scales $\murp,\muhp$ and $\murs,\muhs$ cancels (up to NNLO terms) in
the point-like and hadronic components separately, the dependence upon
$\mug$ in the point-like component is compensated by a corresponding
dependence in the hadronic component.  We finally point out that
equations completely similar to eqs.~(\ref{sigmagp})--(\ref{hadrcomp})
also hold for unpolarized scattering.

In what follows, we will present phenomenological predictions relevant
to HERA physics. To this purpose, we will evaluate both the polarized and
unpolarized cross sections. Since one of the major goals of high-energy
colliders in the polarized mode will be that of studying the polarized
parton densities, we will make a definite choice for the unpolarized
densities throughout the paper, and study the dependence of our predictions 
for the asymmetries upon the polarized densities. Namely, we will adopt
the set MRSA$^\prime$~\cite{MRSAP} for the unpolarized proton,
and the set GRV-HO~\cite{GRVHO} for the unpolarized photon. The value
of $\Lambda_{\sss QCD}$ associated to MRSA$^\prime$ 
($\Lambda_5^{\overline{\rm\sss MS}}=152$~MeV) is well below
the current world average; however, it is closer to the value of
$\Lambda_{\sss QCD}$ associated to GRV-HO and to all the polarized
sets we are going to use. The use of more modern proton sets would only
affect the absolute normalization of the cross sections, without any 
sizeable change in the shape of the distributions. 

As far as the polarized densities in the proton are concerned, DIS data
available at present leave the gluon density largely unconstrained, while
they determine the quark densities to a reasonable extent. Several
parametrizations are therefore available, whose gluon densities are
very different from each other. We will consider the following
sets: GRSV STD and GRSV MAXG~\cite{grsv}, DSS1, DSS2 and DSS3~\cite{dss},
and GS-C~\cite{gs} (see ref.~\cite{FFSV} for a discussion about this choice
and a comparison between the various sets). GRSV STD will be the default set
when studying the perturbative stability of our results.

The polarized densities in the photon are completely unmeasured so far, 
and models for them have to be invoked. To obtain a realistic estimate for
the theoretical uncertainties due to these densities, 
we use the two very different scenarios considered in
ref.~\cite{svgamma}, assuming ``maximal'' ($\Delta f^{(\gamma)} (x,\mu_I^2)=
f^{(\gamma)} (x,\mu_I^2)$) or ``minimal'' ($\Delta f^{(\gamma)} (x,\mu_I^2)
= 0 $) saturation of the positivity constraint\footnote{Strictly speaking, 
this bound is violated in QCD, and the deviations may become sizeable
at small scales. For a discussion on this point, see for example
ref.~\cite{FAR}} $|\Delta f^{(\gamma)}| \le f^{(\gamma)}$ at the input 
scale $\mu_I$ (${\cal{O}}(0.5$ GeV)) for the QCD evolution (here, the 
unpolarized densities are those of the set GRV-HO). The corresponding sets
are denoted as SV~MAX~$\gamma$ and SV~MIN~$\gamma$ respectively.
The former will be our default choice when studying perturbative stability.

In order to compute the NLO results consistently, we will use the
NLO-evolved version of the previously mentioned parton
distributions\footnote{In fact, for most of the sets considered here, there
exists a LO-evolved version as well.}.  We point out that these NLO-evolved
sets will also be used to compute the Born results; indeed, as observed in
ref.~\cite{FFSV}, the use of LO-evolved parton distributions could
introduce differences between the LO and NLO cross sections that are
much larger than those that can be expected on the basis of
perturbative considerations.

Since this paper presents the first NLO calculation of jet cross sections
in polarized photon--hadron collisions, it is mandatory to assess the
effect of the radiative corrections. Lacking an NNLO computation, the most
reliable estimate of the theoretical uncertainty affecting our results
comes from the study of the dependence upon the mass scales entering
eqs.~(\ref{pointcomp}) and~(\ref{hadrcomp}), at fixed values of
the other input parameters. Unless otherwise specified, we will
set all the scales equal to a common value $\mu$, and vary $\mu$
in the range \mbox{$\muo/2\le\mu\le 2\muo$}. Here $\muo$ is the default
scale, which we set equal to half of the total transverse energy
of the event; other sensible choices for $\muo$ give completely
similar results. From eq.~(\ref{sigeppol}), it is clear that
all the dynamical information on the $ep$ process in the 
Weizs\"acker--Williams approximation are contained in the $\gamma p$
cross section. Therefore, we will study the scale dependence of
our results using monochromatic photon--proton collisions at
$E_{cm}^{(\gamma p)}=268$~GeV; the scale dependence at lower centre-of-mass
energies, and in electron--proton collisions, is comparable or milder.

\begin{figure}
\centerline{
   \epsfig{figure=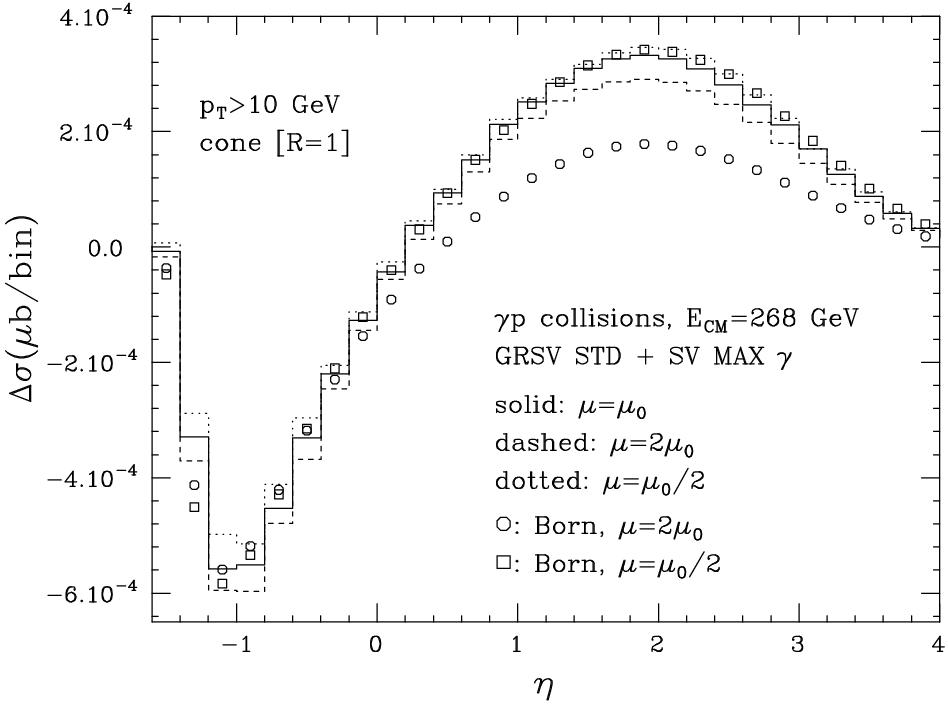,width=0.48\textwidth,clip=}
   \hfill
   \epsfig{figure=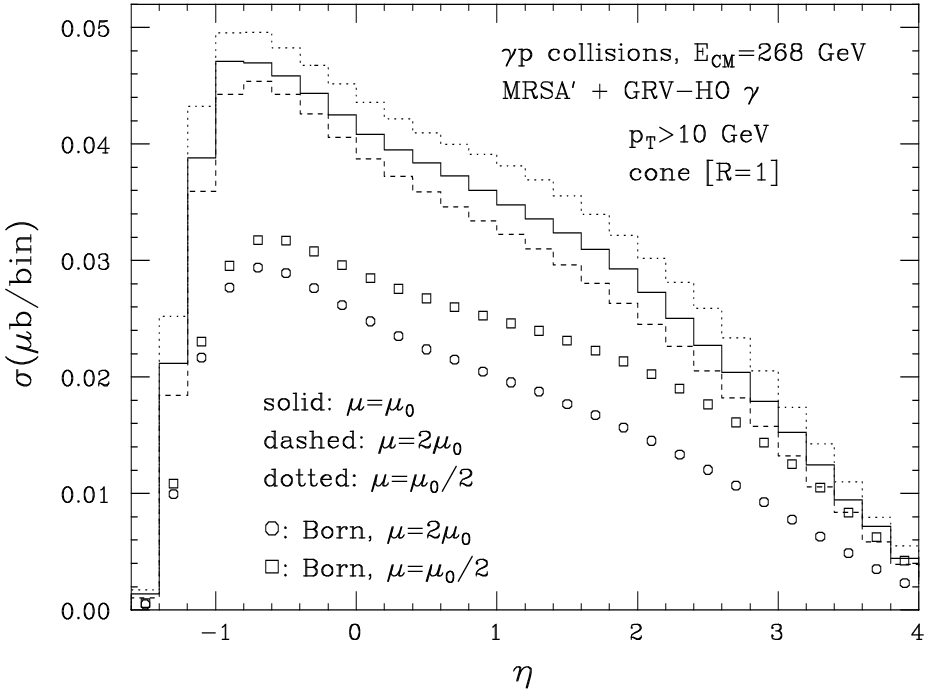,width=0.48\textwidth,clip=} }
\ccaption{}{ \label{fig:eta}
Scale dependence of the single-inclusive jet pseudorapidity in
polarized (left) and unpolarized (right) collisions. 
}
\end{figure}

We start by considering the single-inclusive jet pseudorapidity,
requiring $\pt>10$ GeV, where $\pt$ is the transverse momentum of the jet.
The jet has been defined using the cone algorithm with $R=1$.
The results for polarized and unpolarized cross sections are
presented in fig.~\ref{fig:eta}. The size of the radiative
corrections is larger in the case of unpolarized collisions; in
the case of polarized collisions, the contributions from the
various partonic subprocesses do not have the same sign, and
cancellations occur. 
In both cases, the variation of the scale induces a variation of the
cross section of the order of 10\% over most of the
$\eta$ range considered. The scale dependence is strongly reduced
when going from Born (boxes and circles) to NLO (dashed and dotted
histograms) results. However, this is not true in the region
of negative $\eta$'s, where the scale dependence of the Born
cross section is very small, and smaller than that of the NLO
results. This could be the signal of a failure of the perturbative
expansion, which would be extremely relevant from the phenomenological
point of view, since it is precisely from this region that one
hopes to extract information on the gluon density in the 
proton~\cite{svhera}.
In order to understand this point, we have studied the scale dependence 
of our results more carefully, by varying separately all the
scales that enter eqs.~(\ref{pointcomp}) and~(\ref{hadrcomp}).
At the Born level, the hadronic component has a relatively mild
$\muhs$ dependence, a sizeable $\mug$ dependence, and a very large
$\murs$ dependence. Although the $\mug$ and $\murs$ variations
tend to compensate each other, the latter is dominant, and the 
overall result is a large scale dependence, which is precisely the effect
we see in the positive $\eta$ region in fig.~\ref{fig:eta}, where
the hadronic component is dominant. On the other hand, the point-like
component displays a dependence upon $\muhp$ and $\murp$ of about the same 
size, but opposite in sign (at the Born level, the point-like component 
does not depend upon $\mug$), and a simultaneous variation of the two
scales leads to an almost perfect cancellation of the effects 
induced by them. This is the origin of the small scale dependence 
of the Born result that we see in the negative $\eta$ region in 
fig.~\ref{fig:eta}.
When going to NLO, in the hadronic contribution the $\muhs$ dependence
basically reduces to zero, and the $\murs$ dependence is sizeably
reduced; this is what we expect, as remarked after eq.~(\ref{Deltafdef}).
The $\mug$ dependence is also reduced, although it remains sizeable; indeed, 
the effects of the variation of $\mug$ are only partially cancelled in the 
hadronic component, since the complete cancellation (up to NNLO terms) takes 
place when the point-like component is included. In fact, we can observe 
this cancellation by looking at the $\mug$ dependence of the point-like 
component at NLO, whose size is almost the same as that of the hadronic 
component, but opposite in sign. Finally, the $\muhp$ and $\murp$
dependences of the point-like component are largely reduced with respect
to the corresponding dependences at LO.
In conclusion, the scale dependence of the $\eta$ distribution is
consistent with what we expect from perturbative considerations.
The small scale dependence of the Born result in the negative $\eta$ region
is only due to an incidental cancellation, which takes place when fixing
all the scales to the same value. A more sophisticated treatment of
the scale dependence would result in a scale dependence larger at
the Born level than at the NLO level.

\begin{figure}
\centerline{
   \epsfig{figure=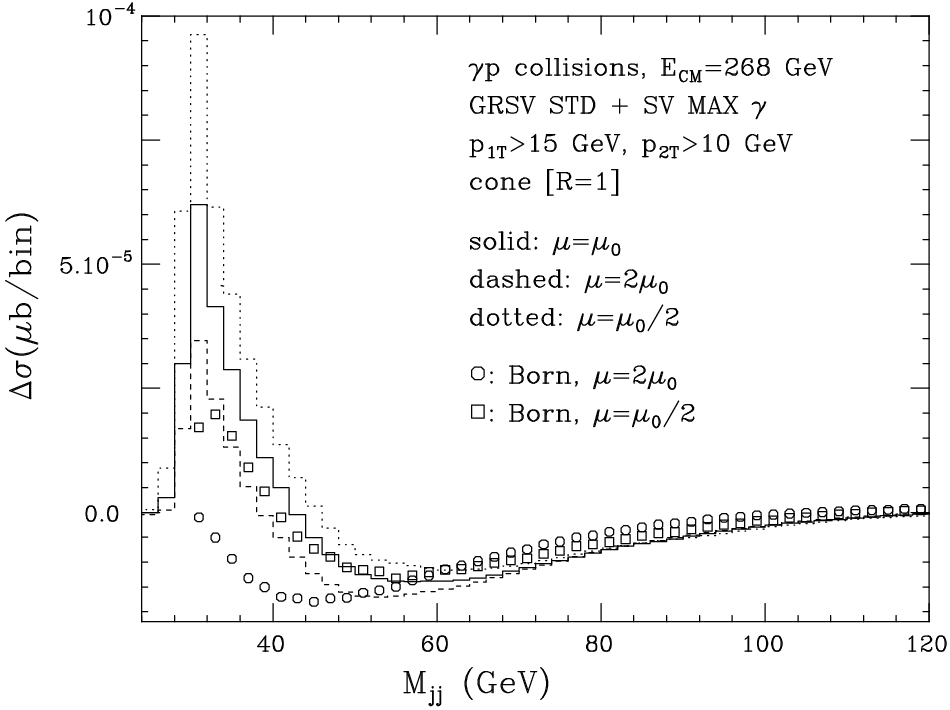,width=0.48\textwidth,clip=}
   \hfill
   \epsfig{figure=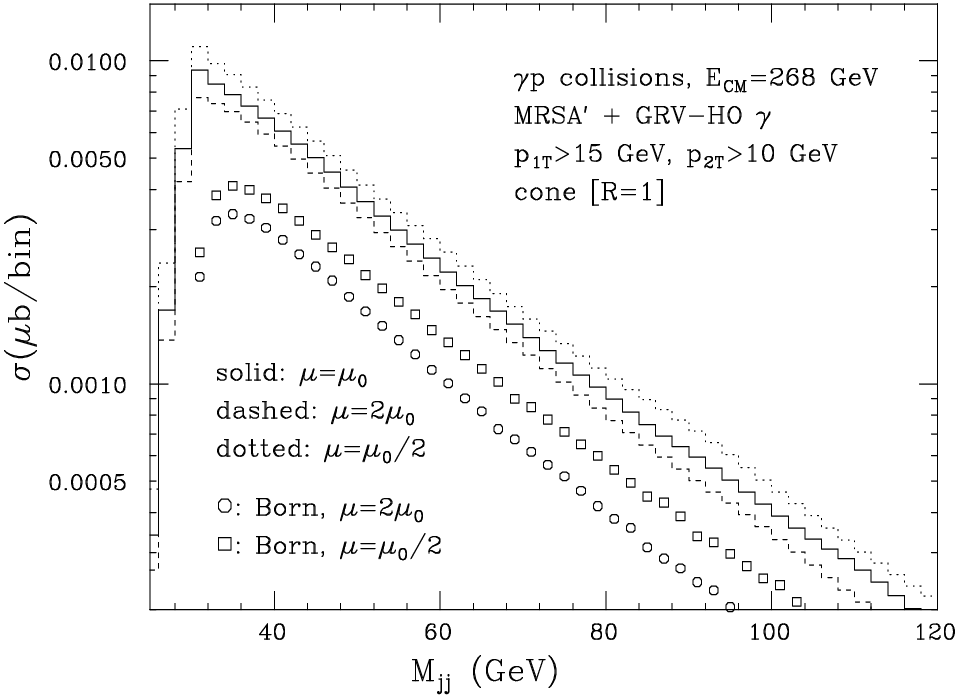,width=0.48\textwidth,clip=} }
\ccaption{}{ \label{fig:mjj}
Scale dependence of the invariant mass of the jet-jet pair in
polarized (left) and unpolarized (right) collisions.
}
\end{figure}                                                              
We also studied several double differential observables, by
considering the two leading jets of each event. As an example,
we present in fig.~\ref{fig:mjj} the invariant mass of the
pair. We require the hardest (next-to-hardest) jet to have 
transverse momentum larger than 15~GeV (10~GeV). We do not
impose any $\eta$ cuts, in order to maximize the scale dependence
(as shown in fig.~\ref{fig:eta}, the scale dependence is
larger for the smallest and largest accessible $\eta$ values,
since they correspond to small transverse momenta). Therefore,
in the case of $\eta$ cuts simulating a realistic geometrical
acceptance, the scale dependence of our results would be smaller.
We see that the case of the invariant mass is pretty similar to
the case of single-inclusive pseudorapidity. The size of
radiative corrections is larger in the case of unpolarized
scattering, and the scale dependence of the polarized result
is comparable to that of the unpolarized one. A reduction of
the scale dependence can be observed when going from LO to NLO,
except for the region close to the threshold, where there is no
contribution at the Born level (due to the asymmetric 
transverse-momentum cuts), and the NLO result is effectively a LO one.

For all the single-inclusive and double-differential jet
observables that we have studied, the same pattern as outlined
above is reproduced. The size of radiative corrections is
usually smaller in the polarized case than in the unpolarized
case. The inclusion of NLO terms reduces the magnitude
of the scale dependence with respect to the LO results, in both
the polarized and unpolarized cases. When considering double-differential 
cross sections, special care has to be taken in
the case where the cuts on the transverse momenta of the two 
leading jets are set to the same value, since there are 
regions in the phase space where the perturbative computation
fails to give a sensible prediction. This situation has been
discussed at length in ref.~\cite{sg2} for unpolarized
jet production. We did not find any major difference in the
case of polarized scattering. We also investigated the effect of 
changing the jet definition, by considering the $\kt$ algorithm
proposed in ref.~\cite{ESalg}. This basically only amounts to a
change in the normalization, while the effects due to the variation 
of the scales are of the same size as those studied above.
Finally, the perturbative stability of the results obtained with
different parton densities is identical to the one presented here.

\begin{figure}
\centerline{
   \epsfig{figure=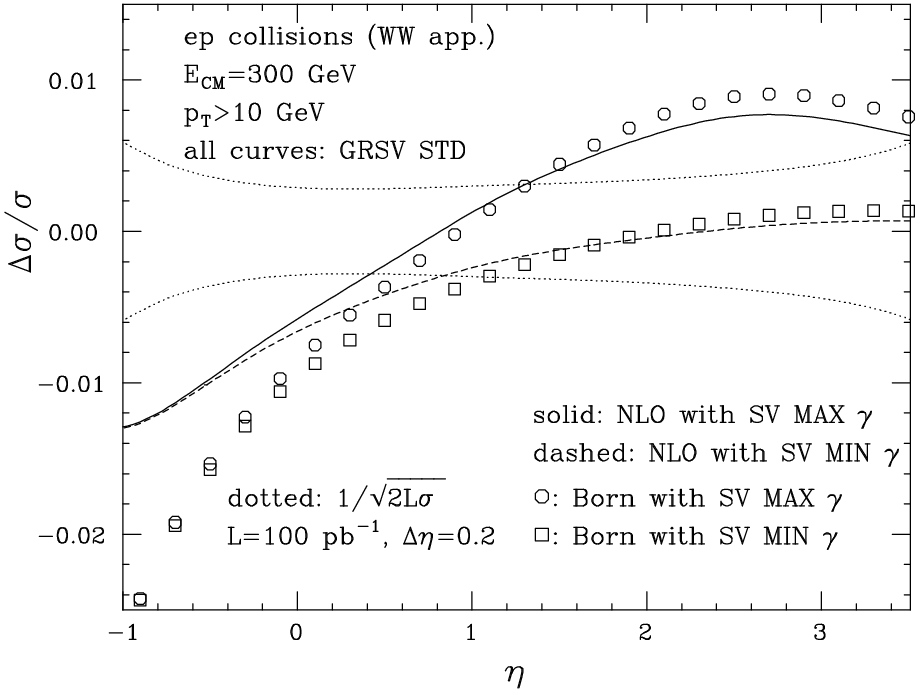,width=0.48\textwidth,clip=}
   \hfill
   \epsfig{figure=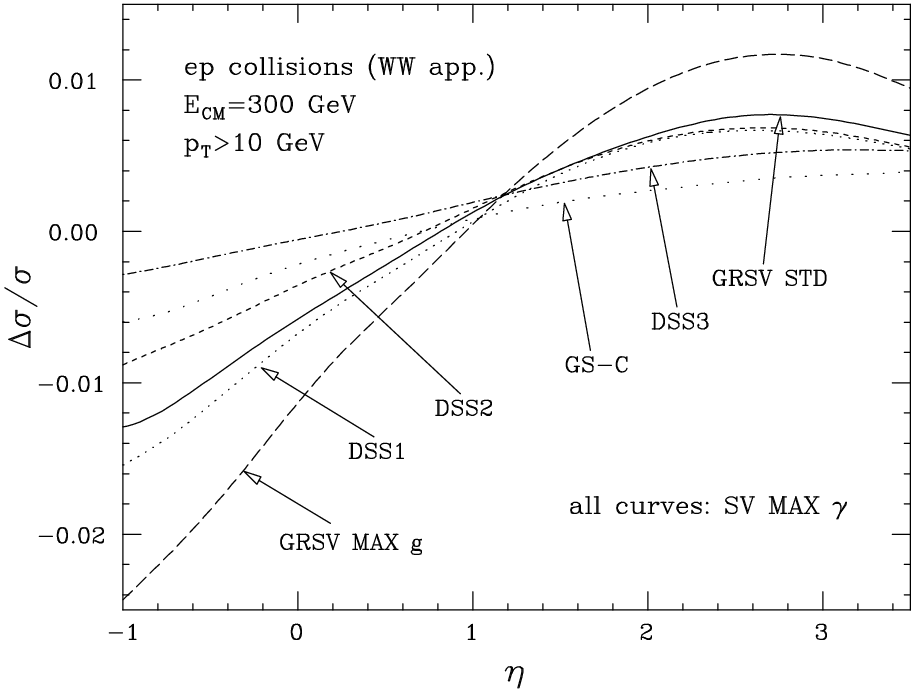,width=0.48\textwidth,clip=} }
\ccaption{}{ \label{fig:asy}
Asymmetries versus pseudorapidity in single-inclusive jet production,
for electron--proton collisions at $E_{cm}=300$~GeV. The predictions for 
several photon (left) and proton (right) densities are shown.
}
\end{figure}                                                              
We now turn to the problem of studying the dependence of our
results upon the available proton and photon polarized parton
densities. To this end, we will consider electron--proton
collisions in the Weizs\"acker--Williams approximation,
setting $E_{cm}^{(ep)}=300$~GeV, $y_{min}=0.2$, $y_{max}=0.8$, and 
$Q^2=0.01$~GeV$^2$ (see eqs.~(\ref{sigeppol}), (\ref{wwfununp}) 
and~(\ref{wwfunpol})). All the scales will be set to the default
value. In fig.~\ref{fig:asy} we present the results for the asymmetry
\beq
{\cal A}_{\eta}=\frac{d\Delta\sigma_{ep}/d\eta}{d\sigma_{ep}/d\eta},
\label{asyeta}
\eeq
where $\eta$ is the pseudorapidity of the single-inclusive jet.
A cut of $\pt>10$~GeV has been applied. On the LHS of the figure,
we have chosen GRSV STD as the polarized proton set, and we evaluated
the results for both the polarized photon sets we consider in this
paper. The Born and NLO results are both shown. We can see that
in the large (positive) $\eta$ region the difference induced by
the choice of the two photon sets is extremely large. On the
other hand, towards negative $\eta$ values this difference
tends to vanish. This is because in that region the point-like
component, which does not depend upon photon densities, is
the dominant one. We can also observe that in the positive $\eta$
region there is a very small difference between the NLO and
LO results, while for negative $\eta$'s the radiative corrections
are positive and reduce the asymmetry considerably. It is worth 
noticing that, as can be seen from fig.~\ref{fig:eta}, this reduction 
is mainly due to the increase of the unpolarized cross section at NLO 
with respect to LO, while the polarized result is basically unchanged.
On the RHS of fig.~\ref{fig:asy}, we show the curves obtained
by fixing the polarized photon set to \mbox{SV MAX $\gamma$}, and
by considering the various polarized proton sets. As expected,
the largest differences can be seen at negative $\eta$ values, where
theoretical predictions can vary for about one order of magnitude.
The shape of the asymmetries is quite similar for all the sets,
except for GS-C; this reflects the fact that the shape of the GS-C
gluon is rather different from the shape of the gluons of the
other sets. The same effect has indeed been observed in
polarized $pp$ collisions~\cite{FFSV}.

On the LHS of fig.~\ref{fig:asy} we also show (dotted lines)
an estimate of the minimum value of the asymmetry observable at HERA. 
This is calculated using the following formula
\beq
\left({\cal A}_{\eta}\right)_{min}= \frac{1}{P_e P_p}
\frac{1}{\sqrt{2\sigma_{ep} {\cal L}\ep}},
\label{Amin}
\eeq 
where ${\cal L}$ is the integrated luminosity, $P_{e(p)}$ is the
polarization of the electron (proton) beam, and the factor $\ep\le 1$
accounts for experimental efficiencies; $\sigma_{ep}$ is the unpolarized
cross section integrated over a range in rapidity ($\Delta\eta$), which
has been chosen, for the present case, equal to $0.2$. In fig.~\ref{fig:asy}
we used \mbox{${\cal L}=100$~pb$^{-1}$}, \mbox{$P_e=P_p=1$} and
$\ep=1$. It follows that, if high luminosity will be collected,
it will be possible to get information on the 
polarized parton densities in the proton, given the fact that
realistic values for the polarization of the beams will be 
such that $P_e P_p\simeq 0.5$. Notice that
the cross section corresponding to sets such as DSS3 or GS-C will be
hardly measurable at HERA, regardless of the luminosity collected.
Of course, the bin size $\Delta\eta$ can be enlarged; 
this would result in a smaller observable asymmetry,
at the price of a loss in resolution. As far as the polarized photon 
densities are concerned, if the ``real'' densities are similar to those 
of the set SV~MIN~$\gamma$, it will be extremely hard to even get the 
experimental evidence of a hadronic contribution to the polarized cross 
section. On the other hand, a set like SV~MAX~$\gamma$ appears to give
measurable cross sections, but this conclusion strongly depends upon 
the polarized proton densities. As observed in ref.~\cite{svhera},
sensible information on the polarized photon densities will be obtained 
only after a more precise knowledge of the proton densities will be 
availabe. At this point, we have to comment upon the effect of the 
non-logarithmic term in the polarized Weizs\"acker--Williams function. It 
turns out that, for the single-inclusive distribution we are studying here,
this term reduces the absolute value of the cross section by a factor
of about 6\%. Since the non-logarithmic term in the unpolarized
case also reduces the cross section by a factor of about 7\%~\cite{sg2},
it follows that the asymmetry is to a good extent independent from
the presence of non-logarithmic terms in the Weizs\"acker--Williams
functions. Finally, if a different jet definition is adopted, the
asymmetries change for less than 1\%; this feature has already
been observed in the case of $pp$ collisions~\cite{FFSV}.

In principle, it is conceivable to study asymmetries also for
monochromatic photon--proton collisions. From the experimental
point of view, this would correspond to selecting events with
the Weizs\"acker--Williams $y$ (that is, the fraction of electron
energy carried away by the photon) in a given narrow range (which 
corresponds to taking $y_{min}$ and $y_{max}$ very close to each other 
in eq.~(\ref{sigeppol})).
If we compute the asymmetry of eq.~(\ref{asyeta}) with
\mbox{$(\Delta)\sigma_{\gamma p}$} instead of $(\Delta)\sigma_{ep}$,
the values we obtain are larger than those shown in fig.~\ref{fig:asy}.
For $E_{cm}^{(\gamma p)}=134$~GeV, the asymmetry can be twice 
as large as that obtained using eq.~(\ref{asyeta}).
Unfortunately, this is not sufficient. Indeed, what is actually
measured at HERA is an electron--proton cross section.
Therefore, to get a physically observable quantity one should
multiply the $\gamma p$ asymmetry by the ratio of the integral
of the polarized Weizs\"acker--Williams function in the range
\mbox{$y_{min}\le y\le y_{max}$} over the same integral calculated
for the unpolarized function. This ratio tends to zero for 
\mbox{$y_{min},y_{max}\to 0$} (\mbox{$y_{max}-y_{min}$} is kept
fixed), thus disfavouring the measurement of the asymmetry at small
photon--proton centre-of-mass energies.

In conclusion, we reported the first calculation of jet cross sections
in polarized photon--hadron collisions, which is accurate to NLO in 
perturbative QCD. For all the observables considered, it has been found
that the size of the radiative corrections is moderate, and that the
scale dependence is smaller than that of the LO result.
The inclusion of the NLO terms reduces the size of the asymmetries
in the pseudorapidity region where the contamination from the 
hadronic photon contribution is minimal. In order to get information
on the polarized gluon densities in the proton and in the photon,
a very high luminosity must be collected at HERA. We also calculated
the universal, non-logarithmic term in the polarized Weizs\"acker--Williams
function. Its contribution to the cross sections is small with respect to
the logarithmic term, analogously to what happens in the unpolarized case.

\section*{Acknowledgements}

We warmly acknowledge W.~Vogelsang for providing us with the
NLO-evolved parton densities in the photon, and G.~Ridolfi for
discussions.  This work was supported in part by the EU Fourth
Framework Programme `Training and Mobility of Researchers', Network
`Quantum Chromodynamics and the Deep Structure of Elementary
Particles', contract FMRX-CT98-0194 (DG 12-MIHT).

\end{document}